\documentclass[useAMS,usenatbib]{mn2e}

\usepackage[none]{hyphenat}

\usepackage[dvips]{graphicx}  
\usepackage{color}
\usepackage{aas_macros}  % required to handle the ADS bibtex journal macros

\newcommand{\Msolar}{\mbox{\,$\rm M_{\odot}$}}        % solar mass
\newcommand{\Lsolar}{\mbox{\,$\rm L_{\odot}$}}        % solar luminosity

\newcommand{\Sun}{\ensuremath{\odot}}
\newcommand{\pg}{\rm PG\,1544+488}
\definecolor{tugca}{rgb}{0.3,0.73,0.09}

\title[ Orbital Elements for  \pg]{Spectroscopic Orbital Elements for the Helium-Rich Subdwarf Binary PG\,1544+488}
\author[H.T.\,\c{S}ener \& C.S.\,Jeffery]{H.~T.~\c{S}ener,$^1$\thanks{email:htss@arm.ac.uk} C.~S.~Jeffery$^{1,2}$ \\
  $^1$Armagh Observatory, College Hill, Armagh, BT61 9DG, UK\\
  $^2$Trinity College Dublin, College Green, Dublin 2, Ireland}
  
\begin{document}

\date{Accepted 2014 February 26. Received 2014 February 21; in original form 2014 January 18}

\pagerange{\pageref{firstpage}--\pageref{lastpage}} \pubyear{2014}

\maketitle

\label{firstpage}

\begin{abstract}

{ \pg\ is an exceptional short-period spectroscopic binary containing
{\it two} sub\-dwarf B stars. It is {\it also} exceptional because the surfaces 
of both components are extremely helium-rich. We present a new analysis of spectroscopy of \pg\ obtained with the
William Herschel Tele\-scope. We obtain improved  orbital para\-meters and atmospheric parameters for each
 component. The orbital  period $P=0.496\pm0.002$\,d,  dynamical mass ratio $M_{\rm B}/M_{\rm A}=0.911\pm0.015$, and spectroscopic radius ratio $R_{\rm B}/R_{\rm A}=0.939\pm0.004$ indicate a binary consisting of nearly identical twins. The data are insufficient to distinguish any difference in surface composition between the components, which are slightly metal-poor (1/3 solar) and carbon-rich (0.3\% by number). The latter indicates that the hotter component, at least, has ignited helium. The best theoretical model for the origin of \pg\ is by the ejection of a common envelope from a binary system in which both components are giants with 
helium cores of nearly equal mass. Since precise tuning is necessary to yield
two helium cores of similar masses at the same epoch, the mass ratio places 
very tight constraints on the dimensions of the progenitor system and on the physics of the 
common-envelope ejection mechanism.} \\
\end{abstract}


\section{Introduction}

Hot subdwarfs are faint blue stars located at the blue end of the horizontal branch. They have
masses in the range of $0.4-0.8$M$_ {\Sun}$ and radii of $\sim$0.1R$_{\Sun}$ and burn helium in
their cores.Hot subdwarfs have lost most of their hydrogen envelopes, but most 
retain a very thin hydrogen atmosphere. After helium core burning they become white dwarfs, 
neither ascending the asymptotic giant branch nor ejecting planetary nebulae.

Subdwarf B (sdB) stars typically have masses $\sim 0.5~{\rm M_{\Sun}}$ and very thin hydrogen-rich
envelopes with masses $\le 0.002 {\rm M_{\Sun}}$ \citep{heber86,saffer94}. Approximately 50\% of hot
subdwarf stars are in binary systems with periods shorter than 30 d \citep{heber09}. Their
companions are generally white dwarfs or M dwarfs but there might also be some systems with more massive unseen components \citep{Geier2010}. Nearly all have hydrogen-rich surfaces.

However $\sim 5 \%$ of hot subdwarfs have helium-rich envelopes \citep{green86,ahmad06}. According to \cite{groth85} and \cite{heber09}, He-rich sdO stars have convective atmospheres, while He-poor sdO and sdB stars mostly have radiative atmospheres. In most sdB stars, gravitational settling produces a He-poor atmosphere but this is not viable for sdO stars. \cite{justham11} suggest that there might be some He-sdB and even He-sdO stars produced as a result of He-rich mergers. These stars might be cool enough to be sdB stars but could still have convective atmospheres.

There are two commonly accepted formation channels to form a{\it single He-rich} hot subdwarf star. The first,
and the dominant mechanism is considered to be the white-dwarf merger channel \citep{webbink84,
iben86}. The mass of the resulting star lies in the range 0.4 -- 0.65~M$_\Sun$
\citep{podsi08.asp392}, depending on the mass of the initial white dwarf binary \citep{han02}.
It is argued that this process actually produces the majority of extremely helium-rich
sdO and sdB stars from the merger of two helium white dwarfs \citep{zhang12a}.
 
The other method to form a single subdwarf starts with a star near the tip of the
first giant branch. If an enhanced stellar wind removes the giant envelope, and the remnant core
ignites helium burning, a subdwarf is formed. In this case, the surface composition of
the subdwarf may depend on when helium-core ignition occurs relative to the star leaving the giant
branch \citep{sweigart04}. The later the helium-core flash occurs, the more efficient is the
``flash-mixing" between the helium core and the residual hydrogen envelope.

Remarkably there are two {\it double} He-rich hot subdwarfs: \pg\ \citep{ahmad04b} and HE\,0301--3039
\citep{lisker04}. \pg\ is an exceptional binary in many respects because both of its components are
He-rich sdB stars. Here the question is no longer how to form a helium-rich subdwarf, but how to form \emph{two}
helium-rich subdwarfs. For normal subdwarfs in binary systems, several evolution channels have been
identified \citep{podsi08.asp392}. In the case of \pg\ and HE\,0301--3039, we first require a scenario to form
a close binary containing {\it two} subdwarfs, and secondly a scenario which requires both subdwarfs
to be {\it helium-rich}.

\subsection{PG1544+488}  

\pg\ was first classified by \citet{green80} as a DA white dwarf and subsequently re-classified as a
hot subdwarf of type sdOD - a `cooler subdwarf with ``pure'' HeI absorption in spectra ...'
\citep{green86}. 

\citet{heber88} suggested \pg\ to be the prototype of a new class of He-sdB stars and published its
spectrum, analysing the He I, CII and CIII lines. The derived effective temperature ($T_{\rm eff}$)
and surface gravity ($\log g$) were 31.0\,kK and 5.1\,[cgs] respectively.

\cite{lanz04} performed a spectroscopic analysis using far-ultraviolet (FUV) spectra from FUSE to
establish a formation method for He-sdB stars and concluded that {``\pg\ fully supports the
flash-mixing scenario''}. An interesting result was the measurement of a large rotational velocity
$v \sin i=100~{\rm km\,s^{-1}}$ in \pg, whereas $v \sin i$ was at most $30~{\rm km\,s^{-1}}$ for the other
two He-sdBs observed with FUSE. 

Up to this point, \pg\ had been deemed to be a single star. Indeed, at low resolution it does not
show radial velocity variations. Observing at higher spectral resolution, \citet{ahmad04b}
discovered spectral lines which were split by a varying amount. They subsequently found that the
integrated FUV spectrum showed large velocity shifts when re-extracted into short exposure spectra.
They therefore identified \pg ~as a spectroscopic binary system composed of two low-mass He-sdB type
stars. An initial solution gave an orbital period of 0.482 d and a mass ratio $q=1.7\pm0.2$.

\citet{ahmad04b} argued that \pg\ must have formed as a result of close binary evolution followed by
the ejection of a common envelope(CE). Both stars must have evolved off the main sequence, so that they
both have a helium core, and the system subsequently must have passed through a CE
phase, including CE ejection, just before helium ignition in the more massive
component. Meanwhile, ~\cite{sweigart04} discussed the shallow and deep flash mixing scenarios for
He-rich \pg\ and two other stars. Their conclusion was to suggest a deep flash-mixing scenario for
\pg\ which considered the hydrogen envelope was mixed deeply into the helium flash layer and
therefore the surface was heavily diluted by helium and carbon.  
\cite{justham11} studied the
formation of binary He-rich subdwarf stars evolved through double-core CE evolution.
This results in a close binary system which contains the  exposed cores of 
both original stars. This occurs by the simultaneous ejection of the envelopes of both stars in the
binary, spiralling inwards in an envelope composed by the merging of the components' envelopes.

In view of the extraordinary difficulty of devising an evolutionary scenario that will
result in the formation of {\it two} helium-rich subdwarfs in a short-period binary, it is of utmost
importance to establish the fundamental properties of both components and, in particular, the mass
ratio. In the following paper, we extract the 2005 observations of \pg\ (\S 2) and obtain a new
orbital solution (\S 3). Spectroscopic quantities are derived for each component (\S 4) and
discussed in terms of evolutionary considerations (\S 5). Overall conclusions and questions for
future work are presented in \S 6. 

%% ==  S.2 ===================================================

\begin{figure} 
\centering 
\includegraphics[width=0.47\textwidth]{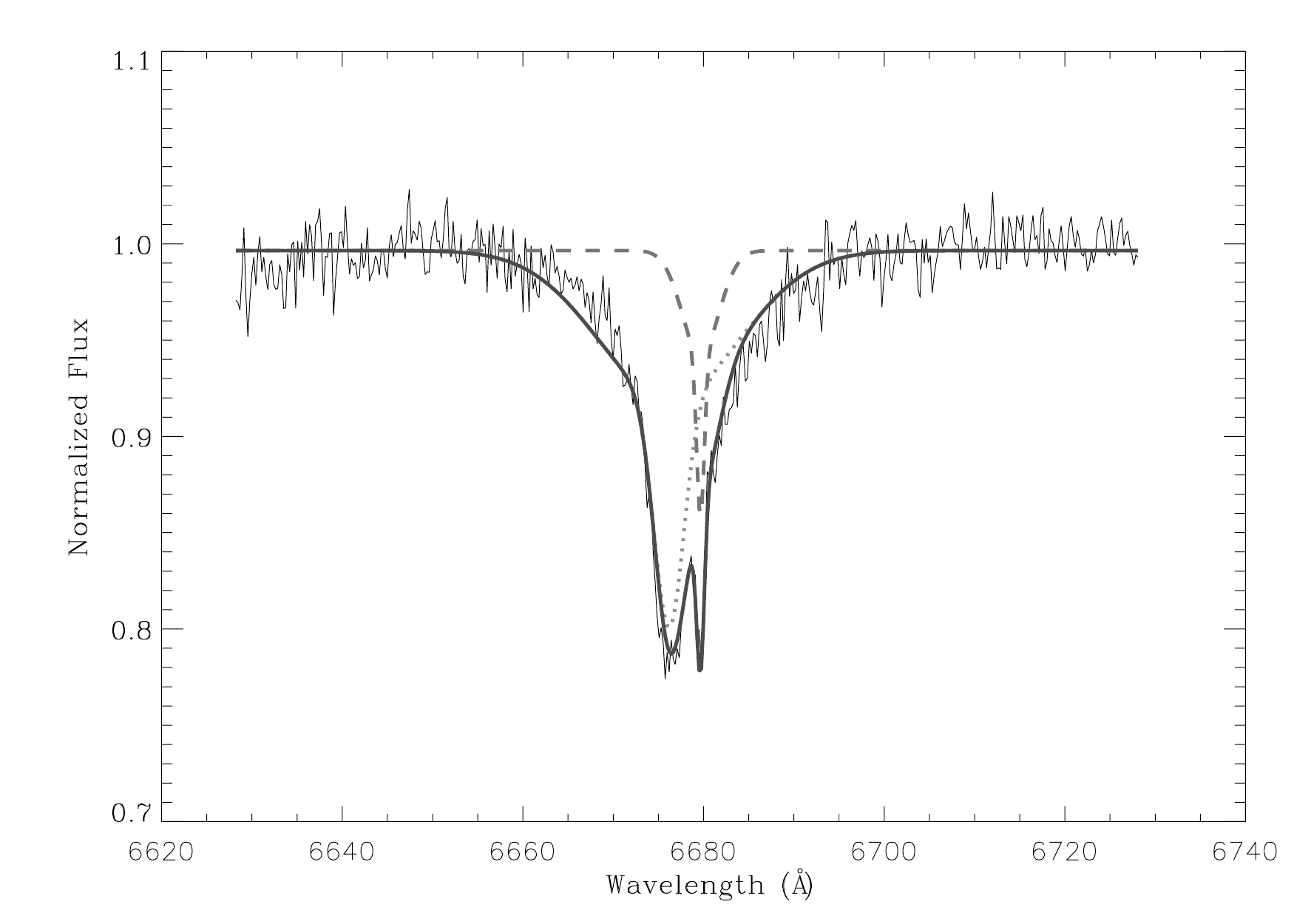}
\caption{A section of red spectrum r746986 showing the He{\sc i} 6678 absorption line, together with
the Gaussian fit used to measure the radial velocities of the components. 
In this example, the stronger blue-shifted component (A: dotted) has a heliocentric radial velocity $-96.0\pm3.1~{\rm km\,s^{-1}}$, 
the weaker red-shifted component (B: dashed)  has a radial velocity  $60.7\pm2.2~{\rm km\,s^{-1}}$. 
The bold line is the combination of both components. }
\label{fig:redfit} 
\end{figure}

\begin{table}
\caption{Observing log showing the WHT/ISIS run number, the heliocentric Julian date for
the midpoint of each observation, orbital phase with the ephemeris of Table {\ref{orbsol}}, and heliocentric
radial velocities and errors for each component. }
\label{observs}
\setlength{\tabcolsep}{5pt}
\begin{center}
{\small
\begin{tabular}{lcccccc}
\hline
Run& Phase&HJD& $V_{\rm A}$ & $\pm$ & $V_{\rm B}$ &  $\pm$ \\
     &&--2453510 & \multicolumn{2}{c}{[km\,s$^{-1}$]} & \multicolumn{2}{c}{[km\,s$^{-1}$]}  \\
\hline
r746618 & 0.1844 & 7.4469&   59.8 &   3.3 & -117.2 &   2.4 \\
r746620 & 0.1983 & 7.4538&   53.3 &   3.3 & -122.3 &   2.4 \\
r746708 & 0.2782 & 7.4934&   61.8 &   3.0 & -115.5 &   2.1 \\
r746710 & 0.2923 & 7.5004&   68.0 &   3.3 & -114.9 &   2.3 \\
r746798 & 0.3707 & 7.5393&   52.1 &  10.2 & -84.2 &   7.3 \\
r746800 & 0.3860 & 7.5469&   43.0 &   3.6 & -79.0 &   2.6 \\
r746877 & 0.4631 & 7.5851&    6.4 &   3.4 & -40.7 &   2.5 \\
r746890 & 0.4784 & 7.5927&    7.1 &   4.1 & -40.1 &   2.9 \\
r746859 & 0.5639 & 7.6351&  -58.9 &   5.2 &   4.6 &   3.7 \\
r746962 & 0.5778 & 7.6420&  -68.8 &   2.7 &  19.4 &   1.9 \\
r746984 & 0.6535 & 7.6795&  -96.7 &   3.0 &  50.3 &   2.1 \\
r746986 & 0.6676 & 7.6865&  -96.0 &   3.1 &  60.7 &   2.2 \\
r747003 & 0.7291 & 7.7170& -101.3 &   3.0 &  69.3 &   2.1 \\
r747006 & 0.7434 & 7.7241& -107.9 &   3.4 &  78.3 &   2.4 \\
r747516 & 0.1226 & 8.4080&   37.1 &   2.8 & -93.0 &   2.0 \\
r747518 & 0.1367 & 8.4150&   42.1 &   3.4 & -99.4 &   2.4 \\
r747611 & 0.2278 & 8.4602&   57.1 &   3.0 & -123.3 &   2.1 \\
r747613 & 0.2434 & 8.4679&   56.3 &   2.8 & -126.5 &   2.0 \\
r747714 & 0.3384 & 8.5150&   63.5 &   4.3 & -83.7 &   3.1 \\
r747735 & 0.4644 & 8.5775&   -4.8 &   4.3 & -49.6 &   3.1 \\
r747737 & 0.4783 & 8.5844&   -7.9 &   8.1 & -43.0 &   5.8 \\
r747742 & 0.6201 & 8.6547&  -87.4 &   3.6 &  42.5 &   2.6 \\
r747745 & 0.6340 & 8.6616&  -87.3 &   2.7 &  49.6 &   1.9 \\
r747839 & 0.7473 & 8.7178& -112.1 &   4.3 &  67.9 &   3.1 \\
r747841 & 0.7612 & 8.7247& -105.0 &   4.2 &  70.3 &   3.0 \\
\hline
\end{tabular}
}
\end{center}
\end{table}

\section{Observations} 
\label{section:observations}

Observations of \pg\ were obtained by Amir Ahmad with the William Herschel Telescope dual beam ISIS
spectrograph on the nights of 2005 May 26 and 27\footnote{A preliminary analysis based on these
observations exists only in abstract form \citep{ahmad07.iaus240}.}. Three wavelength regions centred at 6815\AA\
(red), 4700\AA\ (long blue) and 4151\AA\ (short blue) with a mean dispersion of 0.22\AA\ per channel
were observed approximately once every hour with two exposures each of 600 s. A list of the red observations is shown in Table 1. Data reduction followed standard procedures including
debiasing, cosmic ray removal, flat fielding, sky subtraction, wavelength calibration, scrunching
and normalization using STARLINK packages~(http://starlink.jach.hawaii.edu). At quadrature, both
components are easily resolved (Fig.~\ref{fig:redfit}). A typical line profile shows a strong and a
weak component, which we shall label A and B respectively.

To measure radial velocities from red spectra, the HeI~6678.15\AA~ line was isolated. A baseline and
two Gaussians were fitted to each spectral line using the {\sc IDL} procedure {\sc mpfit} \citep{mpfit}.
Gaussian widths and intensity ratios were fixed in order to keep consistency between the fits to each
spectrum. Radial velocities were directly calculated from the centroids of each component. Fig.
\ref{fig:redfit} shows a sample spectrum and the Gaussian fits to components A and B which led us to determine the radial velocities of the components {as given in Table \ref{observs}.}

Due to the  lower resolution of the blue spectra, the line  splitting and  shifts are harder to measure.
Therefore computing and measuring radial velocities from the cross-correlation
function was preferred to fitting Gaussians to the line profiles. A template spectrum was produced
by calculating the mean of four spectra close to conjunction given in Table~{\ref{bluespec}}. 
A mask was applied to exclude regions between 4500\AA~-- 4670\AA~and 4730\AA~-- 4840\AA~in a way to include strong lines. Each spectrum was cross-correlated with the template. The
radial velocity shifts were measured by fitting a parabola to the peaks of the resulting
cross-correlation function. To measure the template velocity, the template spectrum was
cross-correlated with a local thermodynamic equilibrium (LTE) model spectrum of $T_{\rm eff}=30\,000 {\rm K}$, $\log g = 5.0$ and 
helium abundance $n_{\rm He}=0.99$ \citep{behara06}. Only velocities obtained close to quadrature
were deemed reliable, and these were used only in the spectroscopic analysis discussed in \S\,4.

%% ==  S.3 ===================================================

\begin{figure}
	\centering
		\includegraphics[width=0.47\textwidth]{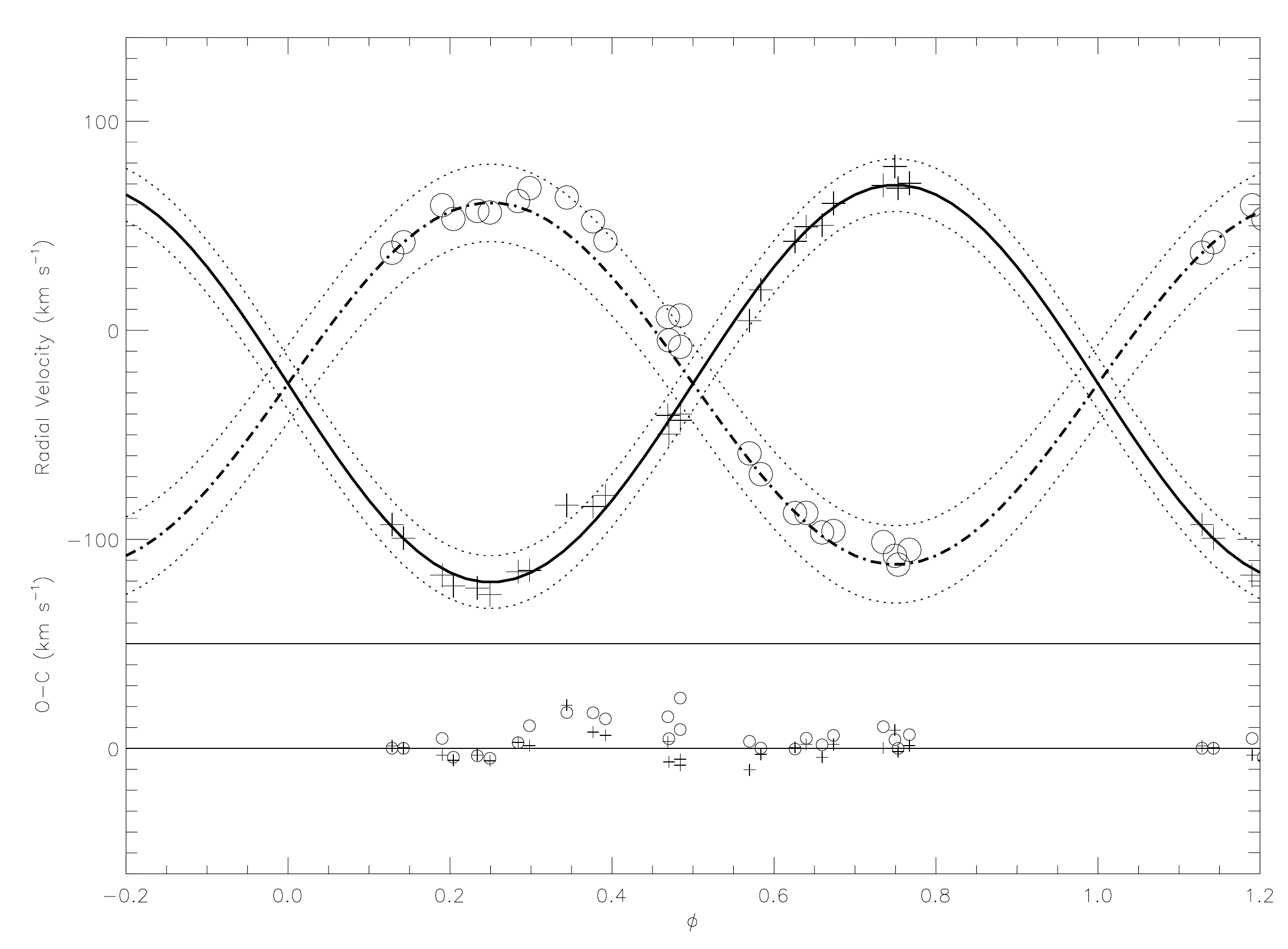}
\caption{The orbital solution for \pg\ obtained using the 2005 He{\sc i}\,6678 observations. Plus signs represent the primary component, open circles represent the secondary. The top panel shows the orbital solution and $\pm 2 \sigma$  envelope. The bottom panel shows the difference between solution and observation, using the same symbols.  }
	\label{fig:rvcurve}
\end{figure}

\begin{table}
\caption{Orbital solution for PG1544+488.}
\label{orbsol}
\smallskip
\setlength{\tabcolsep}{5pt}
\begin{center}
{\small
\begin{tabular}{ccc}
\hline
                  & This study & \cite{ahmad04b}  \\
\hline
$T_0$ [HJD]                      &$53517.3538 \pm0.0099$ &          \\ 
$P$ [d]                          &  $0.496 \pm0.002$ & $0.48\pm0.01$    \\
$\gamma$ [${\rm km\,s^{-1}}$] &  $-25.5\pm0.4$     & $-23\pm4$\\
$K_{\rm A}$  [${\rm km\,s^{-1}}$]       &  $86.6\pm0.5$      & $57\pm4$\\ 
$K_{\rm B}$  [${\rm km\,s^{-1}}$]       &  $95.0\pm0.4$       & $97\pm10$\\
\hline
$q \equiv M_{\rm B}/M_{\rm A}$              &  $0.911\pm0.015$   &  $0.59\pm0.07^{\ast}$\\
$a \sin i$ [R$_\Sun$]      &$1.78\pm0.011$           &  $1.47\pm0.13$\\
$M_A \sin^3 i$ [M$_\Sun$] &   $0.161\pm0.002 $  &  $0.114\pm0.021$\\
$M_B \sin^3 i$ [M$_\Sun$] &   $0.147\pm0.002$  &  $0.067\pm0.011$\\
\hline
\multicolumn{3}{l}{ $\ast$ \citet{ahmad04b} gave $q=M_{\rm A}/M_{\rm B}$. }
\end{tabular}
}
\end{center}
\end{table}

\section{Orbital Solution}

Assuming a circular orbit, the radial velocity curve is solved by fitting a sinusoid plus a 
constant in the form

\begin{equation}
\label{radvelbin}
V_{\rm rad}=\gamma + K \sin{(2\pi f (t-T_0))} 
\end{equation}

\noindent where $K=(2\pi a\sin i)/P$ is the semi-amplitude of the velocity curve, $\gamma$ is the
systemic velocity of the system, $a$ is the semi-major axis of the orbit, $P$ is the orbital period
and the epoch $T_0$ is a time of conjunction. The method used a gradient-expansion algorithm to
compute a weighted least-squares-fit to the observations (IDL function {\sc curvefit}); weights were
the inverse of the measurement errors. The quality of the observational data is such that we have assumed a circular orbit.

{Orbital solutions which included measurements from the blue spectra and radial velocities from the
analysis of \citet{ahmad04b} were attempted but, these seriously degraded the quality of the fit.
Consequently, the adopted orbital solution is that obtained solely from measurements of the
He{\sc i}\,6678 line shown in Table~\ref{observs} and is shown in Table~\ref{orbsol}.  
The best-fitting period is $0.496\pm0.002$\,d.}

The errors on the orbital elements reported in Table~\ref{orbsol} are remarkably small. We have
tested their validity in a number of ways. The errors on the velocity semi-amplitudes and systemic
velocity were verified by (a) increasing the values of the measurement errors by a factor 3, and
(b) splitting the observations into two sets (odd and even data). Test (a) increases the resultant
errors by a factor $<1.3$. Test (b) gave half ranges in $K_{\rm B}$ and $K_{\rm A}$ which were 2.5
and 1.5~${\rm km\,s^{-1}}$, or 6 and 4 $\sigma$, respectively. We suspect the errors on $K_{\rm B}$
and $K_{\rm A}$ are too small by factors of 3 and 2 respectively. The half-range in $\gamma$ matches
$\sigma_{\gamma}$ exactly. The formal error in the orbital period given by {\sc curvefit} was just
0.0005\,d. Tests which omitted various numbers of data and the odd-even test both indicated a larger
error $\sigma_P=0.002\,{\rm d}$ should be admitted. Given that the observations cover only just over
three pulsation cycles, the formal error in the epoch $\sigma_{T_0}=0.1\,{\rm d}$ is reasonable.

The new orbital solution is compared with that of \cite{ahmad04b} in Table~\ref{orbsol}. Apart from
the improved precision the only major difference concerns the semi-amplitude of the second
component, which is much better resolved in the 2005 observations. The velocity semi-amplitudes give
a dynamical mass ratio $q_{\rm dyn}~\equiv~M_{\rm B}/M_{\rm A} = 0.911\pm0.015$, if the errors
$\sigma_K$ are increased by a factor 3, $\sigma_q = 0.023$. The mass ratio is much closer to
unity than initially estimated by \citet{ahmad04b}. The higher velocity resolution of the red He{\sc i} 6678 line and improved phase coverage contribute to the much smaller errors on the new solution. The question remains as to
the total mass of either component, since the inclination is not currently available. For either subdwarf to have ignited helium, a minimum mass of some $\sim 0.45 \Msolar$ and hence an inclination of less than $45^{\circ}$ would be required, which is not unreasonable.

%% ==  S.4 ===================================================

\begin{table}
\caption{Observing log showing the WHT/ISIS run number, the heliocentric Julian date for
the midpoint of each observation, orbital phase with the ephemeris of Table {\ref{orbsol}}, and heliocentric
radial velocities and errors for each component. {These data were not used for the orbital solution given in Table \ref{orbsol}. }}
\label{bluespec}
\setlength{\tabcolsep}{5pt}
\begin{center}
{\small
\begin{tabular}{lcccccc}
\hline
Run& Phase&HJD& $V_{\rm A}$ & $\pm$ & $V_{\rm B}$ &  $\pm$ \\
     &&--2453510 & \multicolumn{2}{c}{[km\,s$^{-1}$]} & \multicolumn{2}{c}{[km\,s$^{-1}$]}  \\
\hline
r746617 & 0.3378 & 8.5150 &  -98.2 & 4.4 &   85.3 & 5.3\\
r746619 & 0.3517 & 8.5219 & -112.4 & 5.3 &   75.6 & 3.7\\
r746709 & 0.4414 & 8.5664 & -122.6 & 3.4 &   59.9 & 0.9\\
r746797 & 0.4638 & 8.5775 &  -81.2 & 3.4 &   38.2 & 2.4\\
r746985 & 0.7606 & 8.7247 &   94.3 & 8.3 & -102.0 & 5.0\\
\hline
\end{tabular}
}
\end{center}

\end{table}

\begin{figure*}
	\centering
		\includegraphics[width=0.9\textwidth]{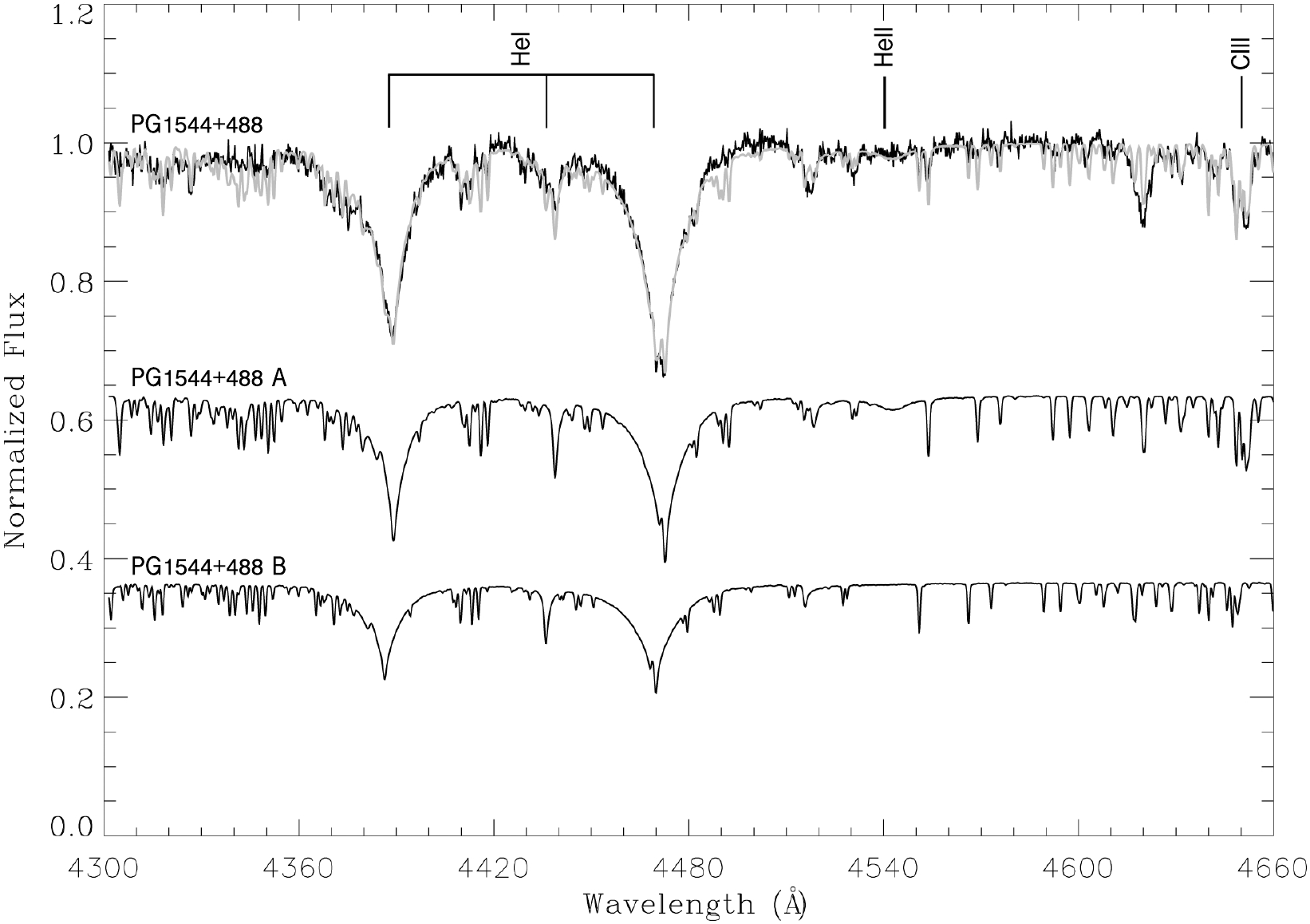}
		\includegraphics[width=0.9\textwidth]{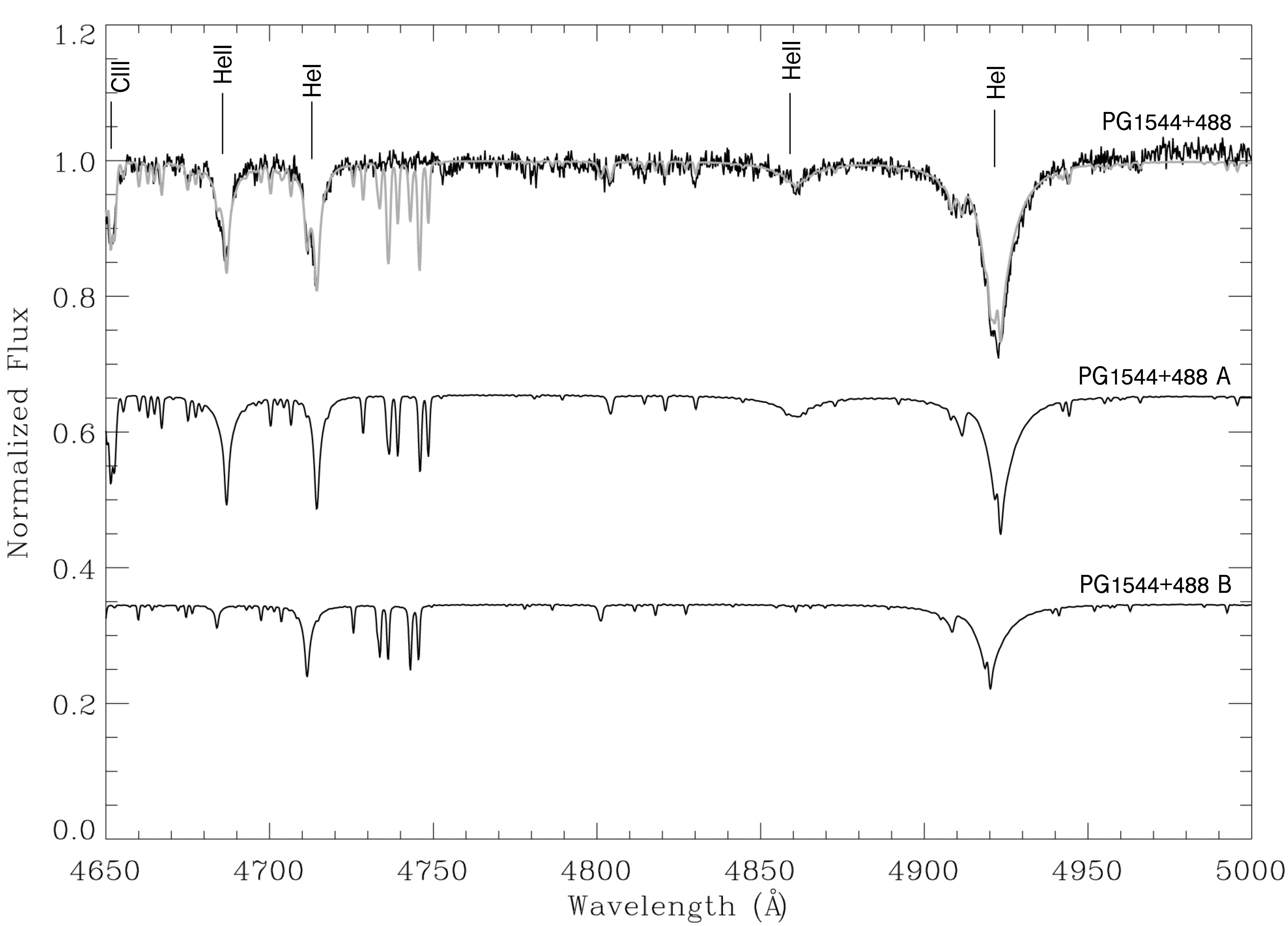}
\caption{ A sample blue spectrum of \pg\ (r676619, Table \ref{bluespec}) (dark line)
is compared with the combined model fit (grey line), and also showing the 
relative model contributions due to the individual components. 
The principal lines due to He{\sc i}, He{\sc ii}, and C{\sc iii} are identified. }  
	\label{fig:4640_5000}
\end{figure*}

\begin{figure}
	\centering
		\includegraphics[width=0.47\textwidth]{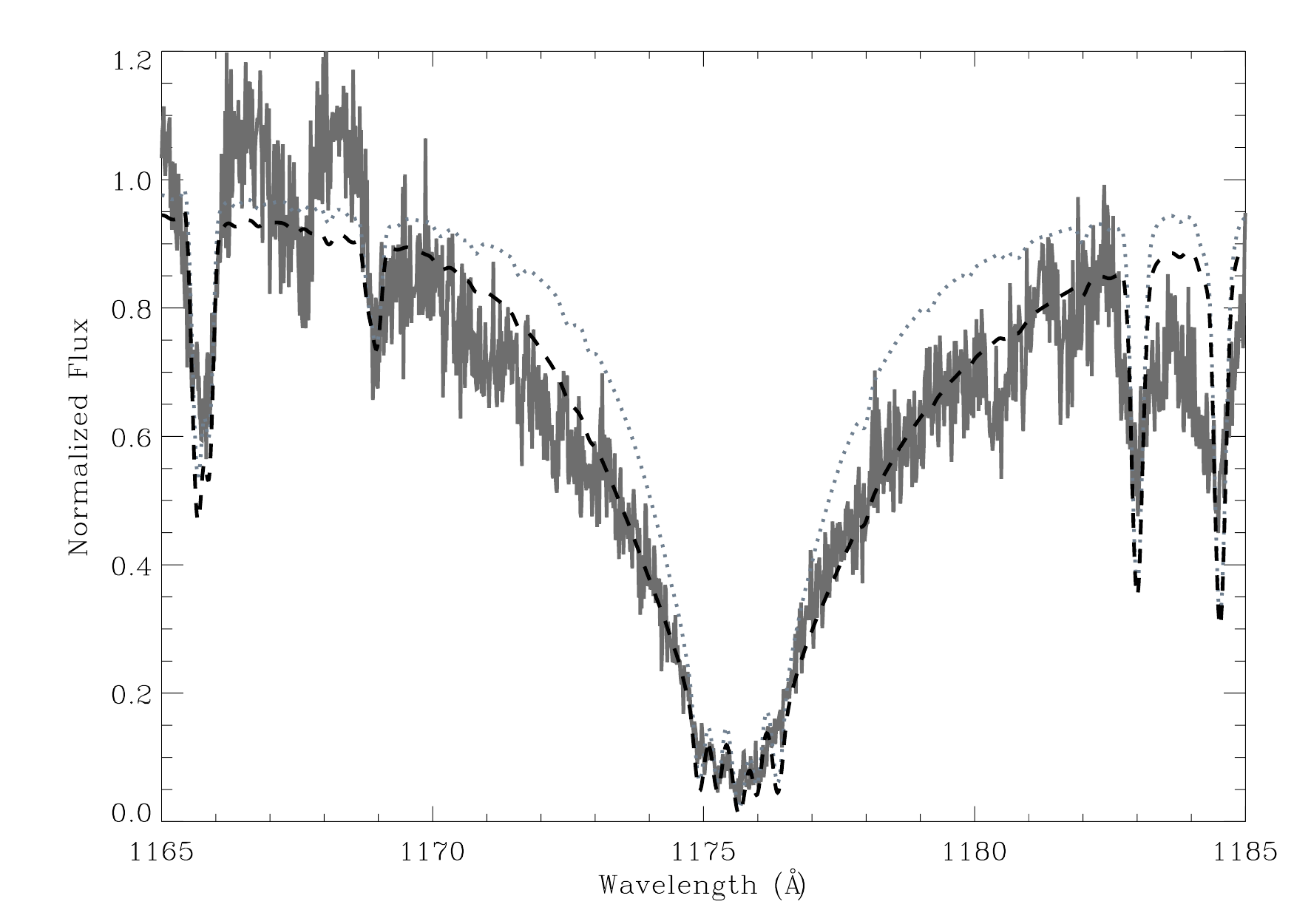}
\caption{ The observed  C{\sc iii} 1176\AA~line profile from FUSE (solid line) 
is compared with theoretical profiles calculated for a metallicity $0.3\times$ solar and for two 
 carbon abundances:  $n_{\rm C}=0.001$ (dotted) and  $n_{\rm C}=0.003$ (dashed). }

	\label{fig:Cline_abund}
\end{figure}

\begin{table}
\caption{Physical parameters of PG1544+488}
\label{physpar}
\smallskip
\begin{center}
{\small
\begin{tabular}{ccc}
\hline
&  A &B \\
\hline
$T_{\rm eff}$ [K]  & $32800\pm270$ & $ 26500 \pm435$ \\
$\log g$ [cgs]  & $5.33\pm0.11$ & $5.35\pm0.15$\\[1mm]
$R_{\rm B}/R_{\rm A}$ & \multicolumn{2}{c}{$0.939\pm0.004$}\\
\hline
$L_{\rm B}/L_{\rm A}$ & \multicolumn{2}{c}{$0.376\pm0.014$}  \\
$q \equiv M_{\rm B}/M_{\rm A}$ & \multicolumn{2}{c}{$0.923\pm0.075$} \\
$\log  L/M$ [\Lsolar/\Msolar] & { $2.10\pm0.11$} & {$1.70\pm0.15$} \\
\hline
\end{tabular}
}
\end{center}
\end{table}

\begin{table}
\caption{Surface Abundances}
\label{abund}
\smallskip
\begin{center}
{\small
\begin{tabular}{ccc}
\hline
Abundance & This study & \cite{lanz04}\\
\hline
$n_{\rm H}$ & $<$0.001&$<$0.002    \\
$n_{\rm He}$ &$0.99\pm0.01$&0.96\\
$n_{\rm C}$ &$0.003\pm0.01$&0.02\\
$[{\rm Fe}]$ & $-0.5\pm0.1$ & -- \\ 
\hline
\end{tabular}
}
\end{center}
\end{table}

\begin{figure}
	\centering
			\includegraphics[width=0.47\textwidth]{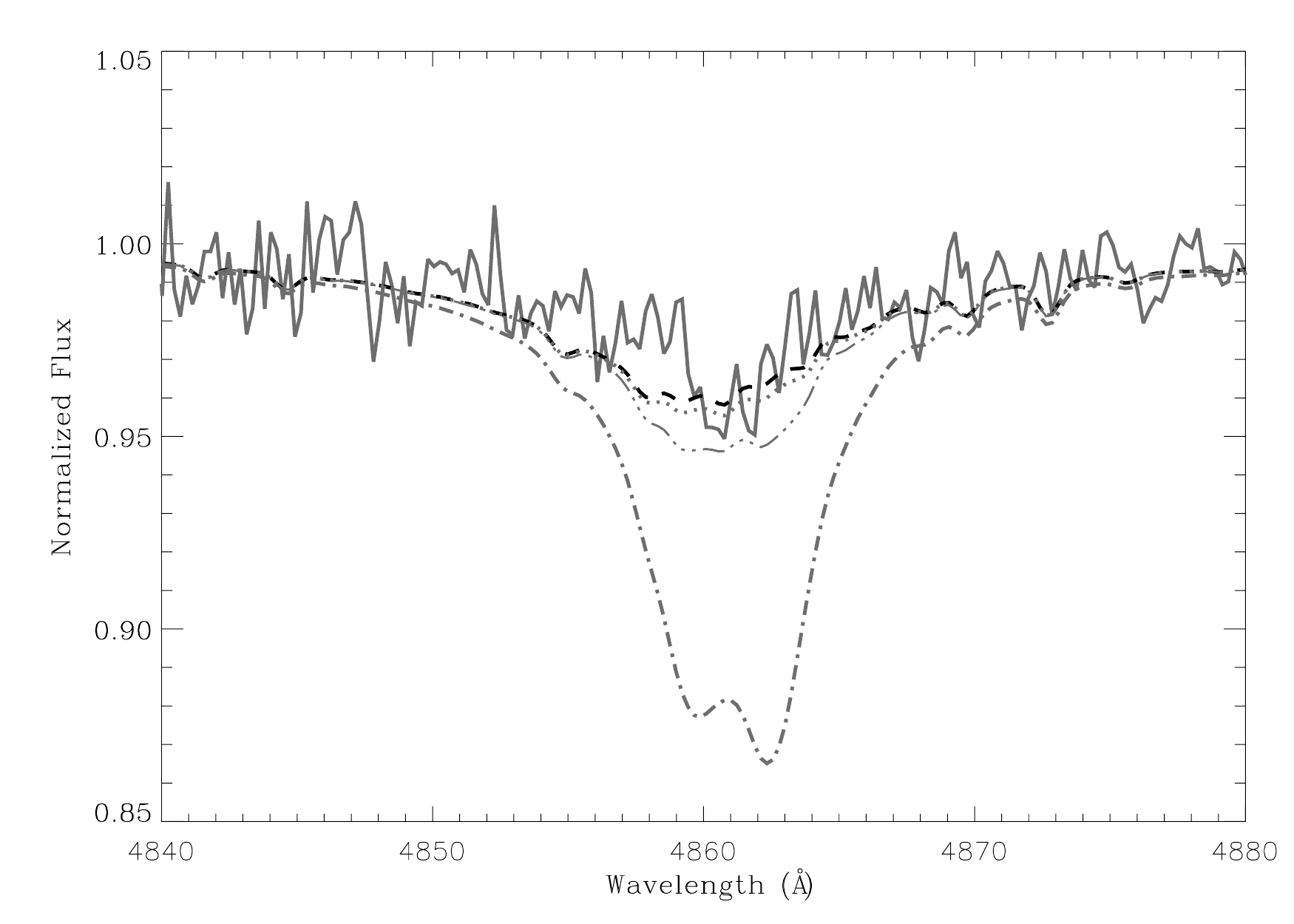}
\caption{ The observed  H$\beta$ 4860\AA~line profile from WHT (solid line: r676619, Table~{\ref{bluespec}}) 
is compared with theoretical profiles calculated for different H abundances:  $n_{\rm H}=0.0001$ (dash), $n_{\rm H}=0.0003$ (dot), $n_{\rm H}=0.001$ (triple dot-dashed) and  $n_{\rm H}=0.001$ (dash-dotted).}

	\label{fig:Hcompare}
\end{figure}

\section{Spectroscopic parameters}

After determining the orbital parameters, we measured the effective temperature
($T_{\rm eff}$) and surface gravity ($\log g$) for each component of PG1544+488.
To do this, we identified the five long-blue spectra closest to quadrature in which
both components could best be resolved. These are indicated in Table~\ref{bluespec} which also shows the velocities obtained by cross-correlation described in \S\ref{section:observations}, and including the  template and heliocentric corrections. 

The observed spectra were then compared with theoretical spectra appropriate for helium-rich
sdB stars. Model atmospheres were calculated with {\sc sterne} \citep{behara06}, which
calculates fully-blanketed plane-parallel atmosphere structures in LTE
for stars with effective temperatures between 10,000 and 35,000 K and any chemical composition.
Emergent spectra are computed with the formal solution code {\sc spectrum} \citep{jeffery01b}. 

Effective temperatures and surface gravities for each component ($T_{\rm A,B}$, $\log g_{\rm A,B}$)
were measured by $\chi^2$ minimization from each spectrum of our five spectra sample. For this, we
used the spectral-fitting software {\sc sfit} \citep{jeffery01b}. This interpolates in a grid of
model spectra to find a best-fitting solution for a single or binary star spectrum. We used model grids
with $4.3<\log T_{\rm eff}<4.7$ and $4<\log g<6.5$ for various chemical 
abundances, {\it i.e.} relative abundances by number of hydrogen $n_{\rm H}$ = [0.0001, 0.001, 0.1],  and carbon $n_{\rm C}$ = [0.001, 0.01, 0.03], and overall metallicities of one and one-third times solar.  In all cases the micro-turbulent velocity was assumed to be 5~${\rm km\,s^{-1}}$ {as a compromise between JL87 \citep{ahmad07} and six other He-sdBs \citep{naslim10}.} Other parameters of the $\chi^2$ solution include the radial velocity, composition and projected rotation velocity $v \sin i$ of each component, as well as the {\it relative} radii of each component $R_i/R_{\rm A}$.

It was not possible to measure a difference between the chemical
composition of the two components. Since the hydrogen was not detected, the adopted solution for both components has relative abundances
(by number) of $n_{\rm H}=0.00001$, $n_{\rm He}=0.99$ and $n_{\rm C}=0.003$. From the weak metal
lines in the spectrum the metallicity is judged to be $\approx0.3$ times the solar value, given in
logarithmic units as [Fe]=--0.5 in Table~\ref{abund}. The results for $T_{\rm eff}$, $\log g$ and
$R_{\rm A}/R_{\rm B}$ are given in Table \ref{physpar}. The composite and constituent theoretical
spectra are shown in Fig.~\ref{fig:4640_5000}; significant lines are labelled including He{\sc i}
4387 and 4471\AA\ which are important in the measurement of surface gravity.

In view of the importance of the carbon abundance to explaining the origin of \pg, we note that
optical lines of C{\sc ii} are notoriously unreliable due to strong departures from LTE. For example,
C{\sc ii} multiplet 1 ($\sim~4730-4750$ ~\AA) is strong in the models, but undetected in the
observations Fig.~\ref{fig:4640_5000}. This is common in the spectra of early-type
hydrogen-deficient stars \citep{jeffery92}. To verify the carbon abundance, we also used the
integrated FUV spectrum observed with FUSE on 2002 March 26. Fig. \ref{fig:Cline_abund} shows the
influence of the carbon abundance on the line wings of C{\sc iii} 1176\AA, which demonstrates that a
model with $n_{\rm C}=0.003$ is substantially better than one with $n_{\rm C}=0.001$. Table
\ref{abund} summarizes our measurements of the chemical composition, and compares them with
\cite{lanz04}.

Since the spectroscopic analysis yields the radius ratio, as well as both temperatures
and gravities, it is trivial to deduce the corresponding luminosity and mass ratios, and also the
luminosity-to-mass ratio for each component (Table \ref{physpar}). In particular, we find the
spectroscopic mass ratio $q_{\rm spec} \equiv M_{\rm B}/M_{\rm A} = 0.923\pm0.075$, in 
excellent agreement with the dynamical mass ratio. Again, {\pg}A is the slightly more massive
component although, in this case, it is not demonstrably different from unity. It is, however,
demonstrably hotter, larger, and more luminous than {\pg}B.

Regarding the surface composition, the hydrogen abundance is below the detection threshold of
$n_{\rm H}< 0.001$. (Fig. {\ref{fig:Hcompare}}). There is no evidence of H$\alpha$,
H$\beta$ or H$\gamma$ distinct from the corresponding He{\sc ii} lines at the same locations. A key
question for interpreting the evolutionary status of \pg\ is whether the surface compositions of the
two stars are identical. This applies in particular to the carbon abundance. The FUV measurement
applies {\it only} to {\pg}A, since the cooler and smaller secondary is invisible at these
wavelengths. The blue spectra are not sufficiently well resolved to measure any difference in this
case, although the equal-abundance fit to C{\sc iii} 4650\AA~ looks good.

%% ==  S.5 ===================================================

\section{Evolutionary Status of \pg}

\begin{figure*}
	\centering
			\includegraphics[width=1\textwidth]{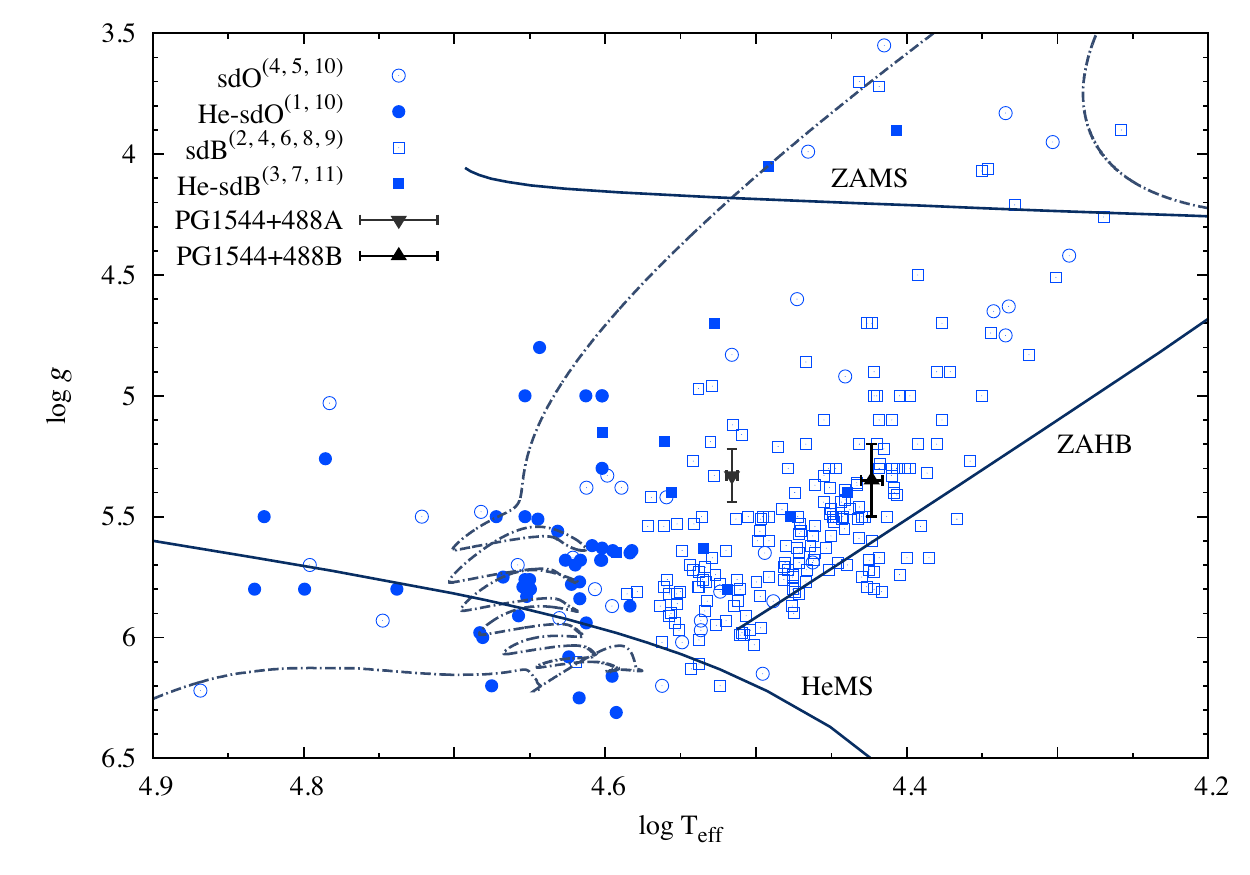}
\caption{The surface gravity -- effective temperature diagram for helium-rich sdB stars
showing the positions of {\pg}A and {\pg}B, other extreme He-rich stars
\citep{ahmad03,stroeer07,naslim10}, theHe-MS, the canonical EHB, and the evolution track for a late hot flasher \citep{bertolami08}. 
References. 1: \citet{Dreizler1990}, 2: \citet{moehler90}, 3: \citet{viton91},
4: \citet{saffer94}, 5: \citet{heber96},6: \citet{moehler97}, 7: \citet{jeffery98},
8: \citet{maxted01}, 9: \citet{lisker04}, 10: \citet{stroeer07}, 11: \citet{naslim10}.
}

	\label{fig:gteff}
\end{figure*}

We have established that the spectroscopic binary \pg\ consists of near-twin extremely helium-rich
and carbon-rich subdwarfs. The effective temperatures and surface gravities are comparable with
other helium-rich subdwarfs (Fig.~\ref{fig:gteff}), \citep{ahmad03,stroeer07,naslim10}. 
In a $g-T_{\rm eff}$ diagram
they are seen to be more luminous than the canonical extended horizontal branch (EHB) for stars with
core masses of $0.47\Msolar$. This is either  because they are more massive than $0.47\Msolar$,
or because they evolving on to or away from the EHB, or on to the helium main sequence (He-MS), or
because they are simply post-giant branch stars evolving through this part of the $g-T_{\rm eff}$
diagram.

It is therefore desirable to establish whether either or both stars are burning helium, and
whether this burning is in the core (He-MS), or in a shell surrounding a degenerate helium core, or
in a shell surrounding a carbon-oxygen core. Neither star lies on the theoretical He-MS, so we exclude 
core helium burning but they are not inconsistent with the pre-white dwarf evolution tracks of very low-mass stars {\citep{driebe99}}.

Since there are no known binaries containing two {\it hydrogen-rich} sdB stars, the
likelihood that we have found such a system in which both components have simultaneously completed
core burning and are evolving away from the EHB and have simultaneously lost their surface hydrogen
seems to be remote. The remaining solution is that both components are helium-shell burning stars
evolving towards the He-MS.

The next question to address is the formation of hydrogen-deficient carbon-rich shell-burning helium
stars. Two major contenders exist.

\citet{brown01}, \citet{lanz04} and \citet{bertolami08} construct models for stars which lose most of their hydrogen-rich
envelope whilst on the giant branch and shortly before core helium ignition. With the hydrogen-shell
extinguished, the degenerate helium core cannot increase in mass, so the star contracts towards the
white dwarf track. However, if core contraction can ignite helium (off-centre) to give a {\it late
helium flash}, the star expands to become a yellow giant and then contracts towards the He-MS. If
the flash is sufficiently late, flash-driven convection will mix material from the ignition shell
through to the stellar surface. Any residual surface hydrogen is ingested (and therefore depleted)
and carbon produced in the helium-shell flash can be mixed to the surface.

Another way to produce extremely helium-rich subdwarfs is to merge two helium white dwarfs
\citep{iben90,saio00}. \citet{zhang12a} showed that double helium white dwarf mergers can be
carbon-rich if the total mass is sufficiently large; the carbon is again produced during helium
ignition and mixed to the surface. This scenario is {\it extremely} unlikely for \pg, since it would
require (a) {\it two} double-white dwarf mergers within a few thousands of years of one another, and
(b) to be preceded by the formation of two compact double-white dwarf binaries in a {\it quadruple}
helium white dwarf system. We reject this as implausible.

Consequently, the late hot flasher model \citet{sweigart04} appears to be preferable. However,
\citet{justham11} point out that the binary nature of \pg\ opens a third possibility. This may occur
if both stars are of sufficiently similar mass that they become red giants with nearly identical
cores {\it and} if the envelope of one star overflows its Roche lobe when the more massive star is
close to helium ignition. In this case, a CE will form which is effectively the
combined envelope of both giants, since both stars overfill their Roche lobes, leading to spiral-in
and, ultimately, ejection of the envelope.

\citet{justham11} point out that double-core CE evolution has not been proven to exist in nature;
\pg\ would be the first such demonstration. Proof of concept might also help to explain the origin
of double neutron star binaries. The parameter space which allows double-core evolution to produce a
double hot subdwarf is small, since the secondary must not only have reached the giant branch by the
time the primary is ready to ignite helium, it must also be able to ignite helium itself.
\citet{justham11} discuss two cases. In one: ``If the secondary is to ignite helium degenerately,
its core mass must be within $\sim5$ per cent of the core mass at the tip of the giant branch when
the envelope is removed \citep{han02}. \ldots The likelihood \ldots is almost negligible''. In the
second: ``stars more massive than \ldots will \ldots ignite helium non-degenerately even if they
lose their envelopes in the Hertzsprung gap. This allows a wider range of parameter space to
potentially produce double-hot-subdwarf systems.'' For ignition of the secondary, the degenerate
case requires the mass ratio of the subdwarfs $q>0.95$, so is marginally consistent with our
measurements but is statistically unlikely. The non-degenerate case is statistically more likely,
with allowed subdwarf mass ratios in the range $0.77<q<1$. In both cases, the mass difference
between the main-sequence progenitors must be less than $\sim0.5$ per cent. A theoretical birth rate
of one double subdwarf for every 3000 hot subdwarfs from both of these channels is consistent with
the uniqueness of \pg\footnote{At present, the other candidate double subdwarf HE0301--3039
\citep{lisker04} remains to be analysed.}. \citet{justham11} also discuss other channels for the
production of dual hot subdwarfs, but state that there is no reason for these to produce stars with
extremely hydrogen-poor surfaces. Regrettably, \citet{justham11} do not indicate what 
surface abundances other than hydrogen and helium might be anomalous in either of the
viable models. The high carbon abundance observed in \pg\ could be produced
in a helium shell flash \citep{bertolami08}, {\it i.e.} following {\it degenerate} double-core evolution; 
the question whether high carbon rules out the non-degenerate case remains to be addressed.

 \section{Conclusions}

The orbital elements of \pg ~have been improved using the best quality data available. The mean
heliocentric velocity of the system ($\gamma$) is -27~${\rm km\,s^{-1}}$, the orbital period is 0.496 d and the mass ratio is close to unity.
 The large radial-velocity amplitude ($K_1+K_2=177$~${\rm km\,s^{-1}}$) allows the spectra of both components to be
resolved and has enabled us to measure the physical properties of both stars independently, despite
their overall similarity. Both components are hydrogen-deficient and helium-rich 
subdwarfs of almost equal mass $(M_{\rm B}/M_{\rm A}=0.911\pm0.015)$ and radius $(R_{\rm B}/R_{\rm A}=0.939\pm0.004)$, with an
orbital separation of a few solar radii. The best chemical composition which matches both optical and
FUV spectra has a metallicity one third of solar and a carbon mass fraction (in the brighter star) of 3 per cent.

\cite{lanz04} demonstrated that the carbon abundance plays an important r\^ole in determining
the best model spectrum for the star, since it dominates the opacity of the atmosphere. 
Our results indicate a lower carbon abundance than \citet{lanz04}, possibly due to other systematic differences 
between the model atmospheres, but still indicate that at least one star has a carbon-enriched atmosphere. 

The results demonstrate that \pg\ consists of near-identical twins, the principal difference being 
in effective temperature with the slightly more massive component being hotter and hence more luminous. 
Since there is negligible surface hydrogen, both stars must have completed main-sequence evolution and
evolved some distance up the giant branch to the point where a CE was formed and 
ejected. 

More detailed work, probably involving observations at higher spectral resolution and signal-to-noise ratio,
will be required to disentangle the spectra of the two components, and to measure their chemical
composition more accurately. In particular, measurements of nitrogen, oxygen and other elements
will be useful for probing the previous evolution. One problem raised is the origin of the
high carbon abundance; is this really a signature of flash mixing produced by degenerate helium-ignition, 
as \citet{lanz04} argue? 
Are both components carbon-rich? Does this give a clue to the {\it actual} masses of these stars, and
also of their progenitors? 
 
At present, \pg\ represents a unique binary system. Further observations may demonstrate that there
are others like it, including HE\,0301-3039 for example. Meanwhile, it remains crucial  for
understanding binary star evolution and it tells us something very important; {\it in a CE binary
containing two helium cores of nearly equal mass -- the entire hydrogen envelopes of both stars are ejected.}
This places a very strong constraint on the physics of the CE ejection mechanism. 

\section*{Acknowledgments}
The results presented in this paper are partly based on grids of model spectra computed using
model-atmosphere codes written by Natalie Behara \citep{behara06}. They are also based on
observations obtained by Amir Ahmad and made with the William Herschel Telescope operated on the
island of La Palma by the Isaac Newton Group in the Spanish Observatorio del Roque de los Muchachos
of the Instituto de Astrofisica de Canarias, and on observations made with the NASA-CNES-CSA Far
Ultraviolet Spectroscopic Explorer. FUSE is operated for NASA by the Johns Hopkins University under
NASA contract NAS5-32985.

The Armagh Observatory is funded by direct grant from the Northern Ireland Department of Culture,
Arts and Leisure.

\bibliographystyle{mn2e}
\bibliography{ehe}

\label{lastpage}

\end{document}